\documentclass[prb,superscriptaddress,showpacs]{revtex4}
\usepackage{graphics}
\usepackage{epsfig}
\newcommand{\om}{\omega}

\newcommand{\E}{{\bf E}}
\newcommand{\kv}{{\bf k}}
\newcommand{\K}{{\bf K}}
\newcommand{\rv}{{\bf r}}
\newcommand{\pv}{{\bf p}}
\newcommand{\R}{{\bf R}}
\newcommand{\Hv}{{\bf H}}
\newcommand{\I}{{\stackrel{\leftrightarrow}{\bf I}}}
\newcommand{\h}{{\stackrel{\leftrightarrow}{\bf h}}}
\newcommand{\green}{{\stackrel{\leftrightarrow}{\bf G}}}
\newcommand{\dyadic}{{\stackrel{\leftrightarrow}{\bf A}}}
\newcommand{\av}{{\bf a}}
\newcommand{\bv}{{\bf b}}
\newcommand{\xv}{{\bf x}}
\newcommand{\ud}{{\bf u}_{d}}

\newcommand{\jv}{{\bf j}}
\newcommand{\mv}{{\bf m}}
\newcommand{\Sig}{{\langle S(\omega) \rangle}}
\newcommand{\ev}{{\bf e}}

\begin{document}

\title{Definition and Measurement of the Local Density of 
Electromagnetic States close to an interface}

\date{\today}

\author{Karl Joulain}
\affiliation{Laboratoire d'\'Etudes Thermiques\\ ENSMA\\ 1 Avenue Cl\'ement Ader
86960 Futuroscope Cedex France}
\affiliation{Laboratoire EM2C\\ Ecole Centrale Paris  CNRS\\ Grande Voie des vignes
 92295 Ch\^atenay-Malabry France}
\author{R\'emi Carminati}
\affiliation{Laboratoire EM2C\\ Ecole Centrale Paris  CNRS\\ Grande Voie des vignes
 92295 Ch\^atenay-Malabry France}
\author{Jean-Philippe Mulet}
\affiliation{Laboratoire EM2C\\ Ecole Centrale Paris  CNRS\\ Grande Voie des vignes
 92295 Ch\^atenay-Malabry France}
\author{Jean-Jacques Greffet}
\affiliation{Laboratoire EM2C\\ Ecole Centrale Paris  CNRS\\ Grande Voie des vignes
 92295 Ch\^atenay-Malabry France}

\begin{abstract}We propose in this article an unambiguous definition of the local density of electromagnetic states (LDOS) in a vacuum near an interface in an equilibrium situation at temperature $T$. We show that the LDOS depends only on the electric field Green function of the system but does not reduce in general to the trace of its imaginary part as often used in the literature. We illustrate this result by a study of the LDOS variations with the distance to an interface and point out deviations from the standard definition. We show nevertheless that this definition remains correct at frequencies close to the material resonances such as surface polaritons. We also study the feasability of detecting such a LDOS with apetureless SNOM techniques. We first show that a thermal near-field emission
spectrum above a sample should be detectable and that this measurement could give access to the electromagnetic LDOS. It is further shown that the apertureless SNOM is the optical analog of the scanning tunneling microscope which is known to detect the electronic LDOS. We also discuss some recent SNOM experiments aimed at detecting the electromagnetic LDOS.

\end{abstract}

\pacs{03.50.De, 07.79.Fc, 44.40.+a, 73.20.Mf }

\maketitle

\section{Introduction}
The density of states (DOS) is a fundamental quantity from which many 
macroscopic quantities can be derived. Indeed, once the  DOS is known, 
the partition function can be computed yielding the free 
energy of the system. It follows that the heat capacity, forces, etc 
can be derived. A well-known example of a macroscopic quantity that 
follows immediately from the knowledge of the electromagnetic DOS is 
the Casimir force\cite{vankampen,gerlach}. Other examples are
shear forces \cite{pendry} and heat transfer \cite{mulet} between two 
semi-infinite dielectrics. Recently, it has been shown that unexpected 
coherence properties of thermal emission at short distances from an 
interface separating vacuum from a polar 
material are due to the contribution to the density of states of 
resonant surface waves \cite{Carminati}. It has also been shown that 
the Casimir force can be interpreted as essentially due to the surface 
waves contribution to the DOS \cite{vankampen,gerlach}. 

Calculating and measuring the local density of states (LDOS) in the 
vicinity of an interface separating a real material from a vacuum
is therefore necessary to understand many problems currently studied. 
The density of 
states is usually derived from the Green function 
of the system by taking the imaginary part of the Green's 
function\cite{Economou,Harrisson}. 
In solid-state physics, the electronic local
density of states at the Fermi energy at the surface of a 
metal can be measured with a Scanning Tunneling Microscope 
(STM)~\cite{Tersoff}. This has been proved by several experiments, in 
particular the so-called quantum corral experiments~\cite{Crommie}. 
Although one can formally generalize the definition of the 
electromagnetic LDOS by using the trace of the imaginary part of the 
Green's tensor\cite{MartinMoreno}, it turns out that this definition
does not yield the correct equilibrium electromagnetic energy density.

Recently, it has been shown theoretically that the STM and the 
Scanning Near-field Optical Microscope (SNOM) have strong 
analogies~\cite{Carminati2}. More precisely, in the weak tip-sample 
coupling limit, it was demonstrated that a unified formalism can be used to relate
the STM signal to the
electronic LDOS and the SNOM signal to electromagnetic LDOS.
SNOM instruments~\cite{SNOM} have been used to perform different 
kinds of
emission spectroscopy, such as luminescence~\cite{Moyer}, Raman
spectroscopy~\cite{Raman} or two-photon fluorescence~\cite{Sanchez}.
For detection of infrared light, apertureless 
techniques~\cite{apertureless}
have shown their reliability for imaging~\cite{Lahrech} as well as
for vibrational spectroscopy on molecules~\cite{Keilman}.
Moreover, recent calculations and experiments have shown that an 
optical analog of the quantum corral could be designed, and that the 
measured SNOM images on such a structure present strong similarities 
with the calculated electromagnetic LDOS~\cite{Colas,Chicanne}. These 
results suggest that the electromagnetic LDOS could be directly 
measured with a SNOM.

The purpose of this article is to show how the electromagnetic LDOS 
can be related to the electric Green-function, and to discuss 
possible measurements of the LDOS in SNOM.
We first introduce a general definition of the electromagnetic LDOS in a vacuum in presence of materials, possibly lossy objects.
 Then, we show that under 
some well-defined circumstances, the LDOS is proportional to the 
imaginary part of the trace of the electrical Green function.
The results are 
illustrated by calculating the LDOS above a metal surface. We show 
next that the signal detected with a SNOM measuring the thermally 
emitted field near a heated body is closely related to the LDOS and 
conclude that the natural experiment to detect the LDOS is to perform 
a near-field  thermal emission spectrum. We discuss the influence of 
the tip shape. We also discuss whether standard SNOM 
measurements using an external illumination can detect the electromagnetic 
LDOS~\cite{Colas,Chicanne}.

\section{Local Density of Electromagnetic States in a vacuum}

As pointed out in the introduction, the LDOS is often defined as being 
the imaginary part of the trace of the electric-field Green dyadic. This 
approach seems to give a correct description in some 
cases~\cite{Colas,Chicanne}, but to our knowledge this definition has 
never been derived properly for electromagnetic fields in a general 
system that includes an arbitrary distribution of matter with 
possible losses. The aim of 
this section is to propose an unambiguous definition of the LDOS.

Let us consider a system at equilibrium temperature $T$.
Using statistical physics, we write the electromagnetic
energy $U(\omega)$ at a 
given \textit{positive} frequency $\omega$, as the product of the DOS by the mean 
energy of a state at temperature $T$, so that
\begin{equation}
    U(\omega)=\rho(\omega) \ \frac{\hbar \omega}{\exp(\hbar
    \omega/k_{B}T) - 1}
    \label{eq:defLDOSbis}
\end{equation}
where $2\pi\hbar$ is Planck's constant and $k_B$ is Boltzmann's 
constant. We can now introduce\cite{Shchegrov}
a \textit{local} density of states by starting with 
the local density of electromagnetic energy energy $U(\rv,\omega)$ 
at a given point $\rv$  in space, and  at a 
given frequency $\omega$. This can be written by definition of the 
LDOS $\rho(\rv,\omega)$ as

\begin{equation}
    U(\rv,\omega)=\rho(\rv,\omega) \ \frac{\hbar \omega}{\exp(\hbar
    \omega/k_{B}T) - 1}
    \label{eq:defLDOS}
\end{equation}

The density of electromagnetic energy is the sum of the electric 
energy and
of the magnetic energy. At equilibrium, it can be calculated using the
system Green's function and the fluctuation-dissipation theorem. 
Let us introduce the electric and magnetic field correlation functions for a stationnary system
\begin{eqnarray}
\mathcal{E}_{ij}(\rv,\rv',t-t') & = &\frac{1}{2\pi}\int d\omega \mathcal{E}_{ij}(\rv,\rv',\omega)
e^{-i\omega(t-t')} =  \left<E_i(\rv,t)E_j^*(\rv',t')\right> \\
\mathcal{H}_{ij}(\rv,\rv',t-t') & = &\frac{1}{2\pi}\int d\omega \mathcal{H}_{ij}(\rv,\rv',\omega)
e^{-i\omega(t-t')}  =  \left<H_i(\rv,t)H_j^*(\rv',t')\right>
\end{eqnarray}
Note that here, the integration over $\omega$ goes from $-\infty$ to $\infty$.
 If $\jv(\rv)$ is the 
electric current density in the system, the electric field reads 
$\E(\rv,\om)=i\mu_0\om\int \green^E(\rv,\rv',\om).\jv(\rv')d^3\rv'$. 
In the same way, the magnetic field is related to the density of 
magnetic currents $\mv(\rv)$ by
 $\Hv(\rv,\om)=\int \green^H(\rv,\rv',\om)\mv(\rv')d^3\rv'$. In these two expressions,
$\green^E$ and $\green^H$ are the dyadic Green functions of the 
electric and magnetic field, respectively. 
The fluctuation-dissipation theorem yields~\cite{Agarwal} that
\begin{eqnarray}
\mathcal{E}_{ij}(\rv,\rv',\omega) & = & \frac{\hbar \omega}{\left[\exp(\hbar
    \omega/k_{B}T) - 1\right]}\frac{\mu_0\omega}{2\pi}Im G^{E}_{ij}(\rv,\rv',\omega) \\
\mathcal{H}_{ij}(\rv,\rv',\omega) & = & \frac{\hbar \omega}{\left[\exp(\hbar
    \omega/k_{B}T) - 1\right]}\frac{\epsilon_0\omega}{2\pi}Im G^{H}_{ij}(\rv,\rv',\omega) 
\end{eqnarray}
If one considers only the positive frequencies $U(\rv,\omega)=4\times\left[\epsilon_0/2\sum_{i=1,3}
\mathcal{E}_{ii}(\rv,\rv,\omega)+\mu_0/2\sum_{i=1,3}
\mathcal{H}_{ii}(\rv,\rv,\omega)\right]$ so that
\begin{equation}
    U(\rv,\omega)=\frac{\hbar \omega}{\left[\exp(\hbar
    \omega/k_{B}T) - 1\right]}\frac{\omega}{\pi c^2}Im 
Tr\left[\green^E(\rv,\rv,\om)+\green^H(\rv,\rv,\om)\right]
    \label{eq:imtr}
\end{equation}
It is 
important to note that the magnetic-field Green function and the 
electric-Green function are not independent. In fact, one has
\begin{equation}
\label{eq:GhvsGE}
\frac{\omega^2}{c^2}\green^H(\rv,\rv',\om)=\left[\nabla_\rv\times\right].\green^E(\rv,\rv',\om).\left[\nabla_{\rv'}\times\right]
\end{equation}
A comparison of Eqs.~(\ref{eq:defLDOS}) and (\ref{eq:imtr}) shows 
that the LDOS of the electromagetic field reads
\begin{equation}
\rho(\rv,\omega)=\frac{\omega}{\pi c^2}Im 
Tr\left[\green^E(\rv,\rv,\om)+\green^H(\rv,\rv,\om)\right]
=f(\green^E)
    \label{eq:ldosimtr}
\end{equation} 
in which $f(\green^E)$ is an operator which will be discussed more 
precisely in the next section.

\section{Discussion}

The goal of this section is to study the LDOS behavior for some well-characterised physical situations, based on the result in 
Eq.~(\ref{eq:ldosimtr}).

\subsection{Vacuum}
In a vacuum, the imaginary
part of the trace of the electric and magnetic field Green functions 
are equal. Indeed, the electric and magnetic 
field Green functions obeys the same equations and have the same 
boundary conditions in this case (radiation condition at infinity). 
In a vacuum, the LDOS is thus obtained by considering the electric 
field contribution only, and multiplying the result by a factor of 
two.
One recovers the familiar result
\begin{equation}
\label{LDOSvac}
\rho_{v}(\rv,\om)=\rho_{v}(\om)=\frac{\om^2}{\pi^2c^3}
\end{equation}
which shows in particular that the LDOS is homogeneous and isotropic.

\subsection{Plane interface}
Let us consider a plane interface separating a vacuum (medium 
1, corresponding to the upper half-space) from a semi-infinite material 
(medium 2, corresponding to the lower half-space) characterised by its complex dielectric 
constant $\epsilon_2(\om)$ (the material is assumed to be  linear, 
isotropic and non-magnetic). Inserting the expressions of the electric 
and magnetic-field Green functions for this geometry \cite{Sipe}
into equation (\ref{eq:ldosimtr}), one finds the expression of the LDOS at a given 
frequency and at a given height $z$ above the interface in vacuum. In 
this situation, the magnetic and electric Green functions are not the 
same. This is due to the boundary counditions at the interface which 
are different for the electric and magnetic fields. In order to 
discuss the origin of the different contributions to the LDOS, we 
define and calculate an electric LDOS ($\rho^E(z,\om)$) due to the 
electric-field Green function only, and a magnetic LDOS 
($\rho^H(z,\om)$) due to the magnetic-field Green function only. The 
total LDOS $\rho^E(z,\om)=\rho^E(z,\om)+\rho^H(z,\om)$ has a clear physical meaning
unlike $\rho^E(z,\om)$ and $\rho^H(z,\om)$. Note that 
$\rho^E(z,\om)$ is the quantity which is
usually calculated and considered to be the true LDOS. In the 
geometry considered here, the expression of the electric LDOS is~\cite{Carsten2}
\begin{eqnarray}
\rho^E(z,\om)&=&\frac{\rho_{v}(\om)}{4}\left\{\int_0^{1}\frac{\kappa 
d\kappa}{p}
\left[2+Re\left(r_{12}^se^{2ip\om 
z/c}\right)+Re\left(r_{12}^pe^{2ip\om z/c}\right)
\left(2\kappa^2-1\right)\right]\right.\nonumber\\
&+&\left.\int_{1}^{\infty}\frac{\kappa d\kappa}{|p|}
\left[Im(r_{12}^s)+\left(2\kappa^2-1\right)Im(r_{12}^p)\right]e^{-2|p|\om 
z/c}\right\}
\label{LDOSelec}
\end{eqnarray}
This expression is actually a summation over all possible plane waves 
with wave number
$\kv=\om/c(\kappa,p)$ where $p=\sqrt{1-\kappa^2}$ if $\kappa\leq 1$ and $p=i\sqrt{\kappa^2-1}$ if $\kappa> 1$. $r_{12}^s$ and 
$r_{12}^p$ are the Fresnel reflection factors between media 1 and 
2 in $s$ and $p$ polarisations, respectively, for a parallel wave 
vector
$\om\kappa/c$ \cite{Carsten}. $0\leq  \kappa\leq 1$ 
corresponds to propagating waves whereas $\kappa> 1$ corresponds 
to evanescent waves. A similar expression for the magnetic LDOS can 
be obtained~:
\begin{eqnarray}
\rho^H(z,\om)&=&\frac{\rho_{v}(\om)}{4}\left\{\int_0^{1}\frac{\kappa 
d\kappa}{p}
\left[2+Re\left(r_{12}^pe^{2ip\om 
z/c}\right)+Re\left(r_{12}^se^{2ip\om z/c}\right)
\left(2\kappa^2-1\right)\right]\right.\nonumber\\
&+&\left.\int_{1}^{\infty}\frac{\kappa d\kappa}{|p|}
\left[Im(r_{12}^p)+\left(2\kappa^2-1\right)Im(r_{12}^s)\right]e^{-2|p|\om 
z/c}\right\}
\label{LDOSmagn}
\end{eqnarray}
Adding the electric and magnetic contributions yields the total LDOS~:
\begin{eqnarray}
\rho(z,\om)&=&\frac{\rho_{v}(\om)}{2}\left\{\int_0^{1}\frac{\kappa 
d\kappa}{p}
\left[2+\kappa^2\left(Re\left(r_{12}^se^{2ip\om 
z/c}\right)+Re\left(r_{12}^pe^{2ip\om z/c}\right)\right)
\right]\right.\nonumber\\
&+&\left.\int_{1}^{\infty}\frac{\kappa^3 d\kappa}{|p|}
\left[Im(r_{12}^s)+Im(r_{12}^p)\right]e^{-2|p|\om z/c}\right\}
\label{LDOS}
\end{eqnarray}
It is important to note that the electric and magnetic LDOS have 
similar expressions, but are in general not equal. The expression of 
$\rho^H(\rv)$ is obtained by exchanging the  
$s$ and $p$ polarisations in the expression of $\rho^E(\rv)$. As a 
result, the two polarisations have
a symmetric role in the expression of the total LDOS $\rho(\rv)$.

The vacuum situation can be recovered from the previous expression 
by setting the values
of the Fresnel reflection factors to zero. The same result is 
also obtained by taking the LDOS at large distance from the 
interface, i.e, for $z\gg \lambda$ where $\lambda=2\pi c/\omega$ is 
the wavelength. 
This means that at large distances, the interface does not perturb the 
density of electromagnetic states.
In fact, $e^{-2|p|\om z/c}$ becomes negligible for the evanescent 
waves
and $e^{2ip\om z/c}$ is a rapidly oscillating function for the 
propagating waves when integrating over $\kappa$. The result is that 
all the terms containing exponential do not contribute to the 
integral giving the LDOS in the vacuum. 

Conversely, at short distance from the interface, $\rho(\rv,\om)$ is 
drastically modified compared to its free-space value. Equations 
(\ref{LDOSelec})-(\ref{LDOS}) show that the Fresnel coefficients and 
therefore the nature of the material play a crucial role in this 
modification. For example, as pointed out by Agarwal \cite{Agarwal}, 
in the case of a perfectly conducting surface, the contribution of 
the electric and magnetic LDOS vanish, except for their  free-space 
contribution. In this particular case, one also retrieves the vacuum 
result. 

We now focus our attention to real materials like metals and 
dielectrics. We first calculate $\rho(\om)$ for aluminum at different 
heights. Aluminum is a metal whose dielectric constant is well 
described by a Drude model for near-UV, visible and 
near-IR frequencies \cite{Palik}:
\begin{equation}
\label{ }
\epsilon(\omega)=1-\frac{\omega_p^2}{\omega(\omega+i\gamma)}
\end{equation}
with $\omega_p=1.747$ 10$^{16}$ rad.s$^{-1}$ and $\gamma=7.596$ 
10$^{13}$ rad.s$^{-1}$.
\begin{figure}
\begin{center}
\includegraphics[width=4in]{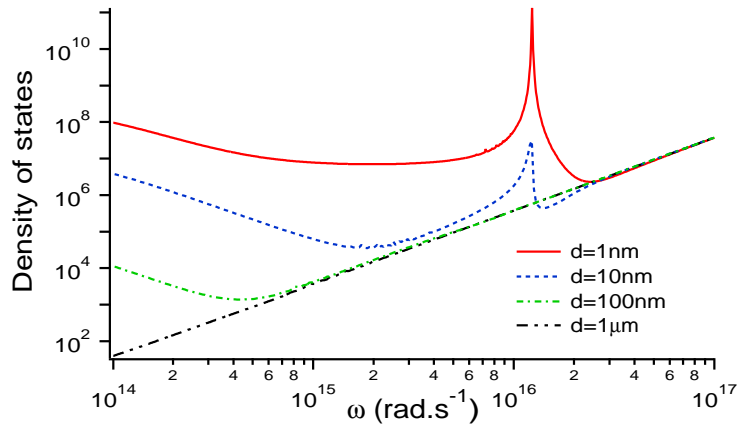}
\caption{LDOS versus frequency at different heights above a 
semi-infinite sample of aluminum.}
\label{ldosz}
\end{center}
\end{figure} 
We plotted in Fig.\ref{ldosz} the LDOS $\rho(\rv,\om)$ in the near 
UV-near IR frequency domain at four different heights. We first note 
that the LDOS increases drastically when the distance to the material 
is reduced. As discussed in the previous paragraph, at large distance 
from the material, one 
retrieves the vacuum density of states. Note that at a given 
distance, it is always possible to find a sufficiently high frequency 
for which the corresponding wavelength is small compared to the 
distance so that a far-field situation is retrieved. When the 
distance to the material is reduced, additional modes are present: 
these are the evanescent modes that are confined close to the 
interface and that cannot be seen in the far field. Moreover, 
aluminum exhibits a resonance around $\omega=\omega_p/\sqrt{2}$. 
Below this frequency, the material supports resonant surface waves 
(surface-plasmon polaritons). Additional modes are therefore seen in 
the near field. This produces an increase of the LDOS close to the interface.
The enhancement is particularly important at the 
resonant frequency  which corresponds to $Re[\epsilon(\omega)]=-1$. 
This behavior is analogous to that 
previously described in Ref.\cite{Shchegrov} for a SiC surface 
supporting surface-phonon polaritons. Also note that in the low 
frequency regime, the LDOS increases. Finally,  Fig.\ref{ldosz} shows 
that it is possible to have a LDOS smaller than that of vacuum at 
some particular distances and frequencies.
\begin{figure}
\begin{center}
\includegraphics[width=4in]{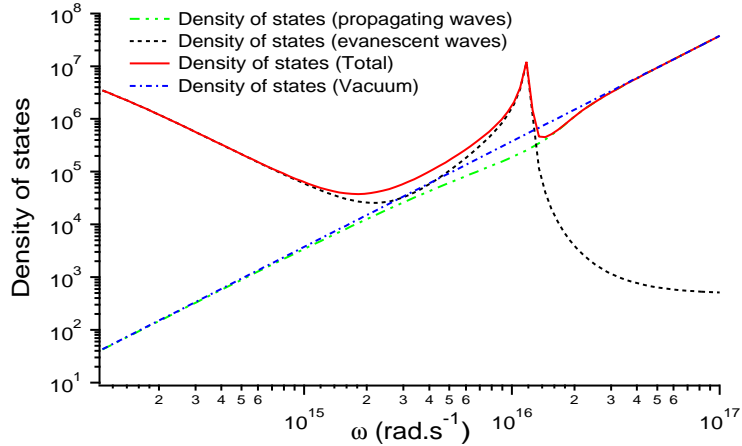}
\caption{Density of states contributions due to the propagating and 
evanescent waves compared to the toatal density of states and the 
vacuum density of states. These quantities are calculated above an 
aluminum sample at a distance of 10 nm.}
\label{fig2}
\end{center}
\end{figure}
Figure \ref{fig2} shows the propagating and evanescent waves 
contributions to the LDOS above an aluminum sample at a distance of 
10 nm. The propagating contribution is very similar to that of the vacuum 
LDOS. As expected, the evanescent contribution dominates at 
low-frequency and around the surface-plasmon polariton resonance), 
where pure near-field contributions dominates.

\begin{figure}
\begin{center}
\includegraphics[width=4in]{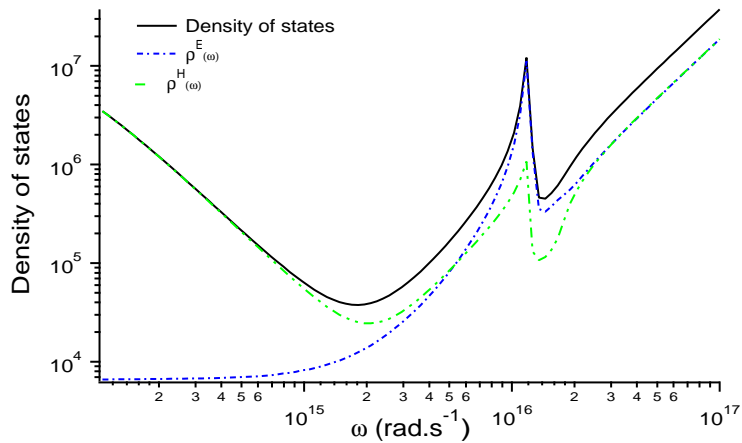}
\caption{LDOS at a distance $z=10nm$ above a semi-infinite aluminum 
sample. Comparison with $\rho^E(\omega)$ and $\rho^H(\omega)$.}
\label{fig3}
\end{center}
\end{figure}

We now turn to the comparison of $\rho(z,\om)$ with the usual 
definition often encountered in the literature, which corresponds to 
$\rho^E(z,\omega)$. We plot in Fig.\ref{fig3} $\rho$, $\rho^E$ and $\rho^H$ above 
an aluminum surface at a distance $z=10 nm$. In 
this figure, it is possible to identify three different domains for 
the LDOS behaviour. We note again that in the far-field situation 
(corresponding here to high frequencies i.e. $\lambda/2\pi\ll z$), the LDOS reduces to the vacuum situation. In this case 
$\rho(z,\omega)=2\rho^E(z,\omega)=2\rho^H(z,\omega)$. Around the 
resonance, the LDOS is dominated by the electric-field Green 
contribution. Conversely, at low frequencies, $\rho^H(z,\om)$ 
dominates. Thus, Fig.\ref{fig3} shows that we have to be very careful 
when using the expression $\rho(z)=\rho^E(z,\omega)$. Above aluminum 
and at a distance $z=10nm$, this approximation is only valid on a 
small range  between $\om=10^{16} rad.s^{-1}$ and $\om=1.5\times
10^{16} rad.s^{-1}$.

\subsection{Asymptotic form of the LDOS in the near-field}

In order to get more physical insight, we have calculated the 
asymptotic LDOS behaviour in the three regimes mentioned above. As we 
have already seen, the far-field regime ($\lambda/2\pi\ll d$) corresponds to the vacuum 
case. To study the near-field situation, we focus on the evanescent 
contribution as suggested by the results in Fig.\ref{fig2}. When 
$\lambda=2\pi c/\omega\gg z$, the exponential term $\exp(-|p|\omega 
z/c)$ is small only for $\kappa\gg\lambda/(4\pi z)\gg 1$. In this 
(quasi-static) limit, the Fresnel reflection factors reduce to
\begin{eqnarray}
\lim_{\kappa\rightarrow\infty}r^s_{12} & = &  \frac{\epsilon 
-1}{4\kappa^2} \\
\lim_{\kappa\rightarrow\infty}r^p_{12} & = &  \frac{\epsilon 
-1}{\epsilon+1}
\end{eqnarray}
Asymptotically, the expressions of  $\rho^E(z,\omega)$ and 
$\rho^H(z,\omega)$ are
\begin{eqnarray}
\rho^E(z,\omega) & = & 
\frac{\rho_v}{|\epsilon+1|^2}\frac{\epsilon^{''}}{4k_0^3z^3} 
\label{asympe}\\
\rho^H(z,\omega) & = & \rho_v\left[\frac{\epsilon^{''}}{16k_0 
z}+\frac{\epsilon^{''}}{4
|\epsilon+1|^2k_0z} \right] \label{asymph}
\end{eqnarray}
At a distance $z=10nm$ above an aluminum surface, these asymptotic 
expressions matches almost perfectly with the full evanescent 
contributions ($\kappa>1$) of $\rho^E$ and $\rho^H$. These 
expressions also show that for a given frequency, one can always find  a distance to the 
interface $z$ below which the dominant contribution to the LDOS will be 
the one due to the imaginary part of the electric-field Green 
function that varies like $(k_0z)^{-3}$. But for aluminum at a distance $z=10nm$, this is not the 
case for all frequencies. As we mentionned before, this is only true 
around the resonance. For example for low frequencies, and for 
$z=10nm$,  the LDOS is actually dominated by $\rho_v\epsilon^{''}/(16k_0 
z)$.

\subsection{Spatial oscillations of the LDOS}
\begin{figure}
\begin{center}
\includegraphics[width=10cm]{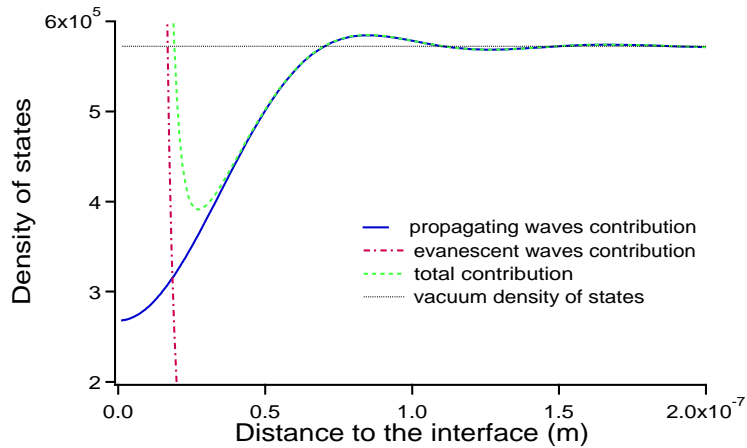}
\caption{LDOS versus the distance $z$ from an aluminum-vacuum 
interface at the aluminum resonant frequency.}
\label{fig4}
\end{center}
\end{figure}

\begin{figure}
\begin{center}
\includegraphics[width=10cm]{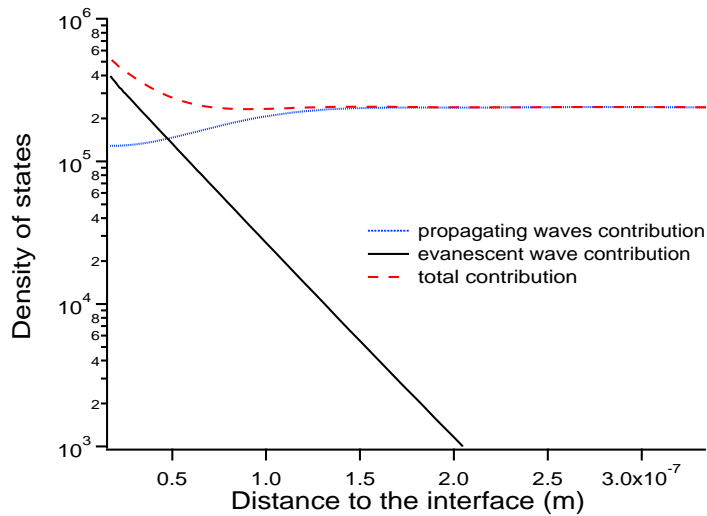}
\caption{LDOS versus the distance $z$ from an aluminum-vacuum 
interface at frequency $\omega=8$ 10$^{15}$ rad.s$^{-1}$.}
\label{fig5}
\end{center}
\end{figure}

Let us now focus on the LDOS variations at a given frequency versus 
the distance $z$ to the interface. There are essentially three regimes. First, for as discussed previously, at distances much larger than the wavelength the LDOS is 
given by the vacuum expression $\rho_v$.
The second regime is observed close to the interface where oscillations are observed. Indeed, at a given frequency, each incident plane wave on the interface can interfere with its reflected counterpart. This generate an interference pattern with a fringe spacing that depends on the angle and the frequency. Upon adding the contributions of all the plane waves over angles, the oscillating structure disappears except close to the interface.
This leads to oscillations around distances on the 
order of the wavelength. This phenomenon is the 
electromagnetic analog of Friedel oscillations that can be 
observed in the electronic density of states near the interfaces~\cite{ashcroft,Harrisson}. As soon as the distance becomes small 
compared to the wavelength, the phase factors $\exp(2ip\om z/c)$ in 
Eq.~(\ref{LDOS}) are equal to unity. For a highly reflecting 
material, the real part of the reflecting coefficients are negative 
so that the LDOS decreases while approaching the surface. These 
two regimes are clearly observed for aluminum in 
Fig.~\ref{fig4}. The third regime is observed at small distances as seen in Fig.~\ref{fig4}. The evanescent 
contribution dominates and ultimately the LDOS always increases as 
$1/z^3$, following the behaviour found in Eq.~(\ref{asympe}). This is 
the usual quasi-static contribution that is always found at short 
distance\cite{Carsten}. 
At a frequency slightly smaller than the resonant frequency, surface 
waves are excited on the surface. These additional modes increase the 
LDOS according to an exponential law as seen in Fig.~\ref{fig5}, a 
behavior which was already found for thermally emitted 
fields~\cite{Carminati,Carsten}. 
At low frequency, the LDOS dependance is given by Eq.(\ref{asymph}). The $1/z$ magnetic term dominates because the $1/|\epsilon+1|^2$ takes large values. The $1/z^3$ contribution equals the $1/z$ contribution for distances much smaller than the nanometer scale, a distance for which the model is no longer valid.

The main results of this section can be summarized as follows.
The LDOS of 
the electromagnetic field can be unambiguously and properly 
defined from the 
local density of electromagnetic energy in a vacuum above a sample at 
temperature $T$ in equilibrium. The LDOS can always be written as a function of the 
electric-field Green function only, but is in general not 
proportionnal to the trace of its imaginary part. An additional term 
proportional to the trace of the imaginary part of the magnetic-field 
Green function is present in the far-field and at low frequencies. At 
short distance from the surface of a material supporting surface 
modes (plasmon or phonon-polaritons), the LDOS presents a resonance 
at frequencies such that $Re[\epsilon(\om)]=-1$. Close to this 
resonance, the  approximation $\rho(z,\omega)=\rho^E(z,\omega)$ 
holds. In the next section, we discuss how the LDOS can be measured.

\section{Measurement of the LDOS}

\subsection{Near-field thermal emission spectroscopy with an 
apertureless SNOM}
In this section we shall consider how the LDOS can be measured  using a SNOM. We consider a frequency range where $\rho$ is dominated by the electric contribution $\rho^E$. We note that for an isotropic dipole, a lifetime measurement yields the LDOS as discussed by Wijnands et al.~\cite{MartinMoreno}. However, if the dipole has a fixed orientation $\xv$, the lifetime is proportionnal to $G_{xx}$ and not to the trace of $\green$. In order to achieve a direct SNOM measurement of the LDOS, we have to fulfill two requirements. First, all the modes must be excited. The simplest way to achieve this is to use the thermally emitted radiation by a body at equilibrium. The second requirement is to have a detector with a flat response to all modes. To analyse this problem we use a formalism recently introduced. 

\begin{figure}
\begin{center}
\includegraphics[width=9cm]{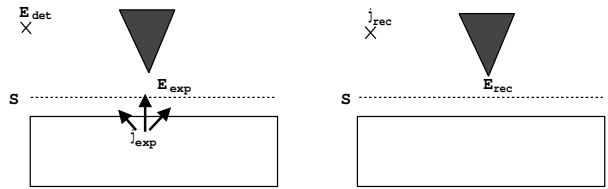}
\caption{Scheme of a scanning near-field optical microscope measuring
a thermally emitted field. (a): experimental situation.
(b): reciprocal (fictitious) situation.}
\label{SNOM}
\end{center}
\end{figure}
We consider a SNOM
working in the detection mode, and detecting the electromagnetic
field thermally emitted by a sample held at a temperature $T$.
The system is depicted in Fig.~\ref{SNOM}. The
microscope tip is scanned at close proximity of the interface
separating the solid body from a vacuum. The signal is measured in 
the far
field,
by a point detector sensitive to the energy flux carried by the 
electromagnetic
field. We assume that an analyzer is placed in front of the detector
(polarized detection). The direction of polarization of the analyzer
is along the direction of the vector $\jv_{rec}$.
If the solid angle $d\Omega$ under which the detector is seen from the
tip is small (a condition we assume for simplicity),
the signal $\Sig$ at the detector, at a given frequency $\omega$, 
reads
\begin{equation}
\Sig =\frac{\epsilon_0 c}{2}\left|\E_d\right(\omega)|^2r^2d\Omega
\label{signal}
\end{equation}
where $\epsilon_0$ is the permittivity of vacuum,
$c$ is the speed of light in vacuum, $r$ is the distance
between the tip and the detector, and $\E_{d}$ is the electric field 
at
the position of the detector.
Let us denote by $\E_{exp}$ (experimental field) the thermal field,
emitted by the sample, in the gap region between the tip and the
sample.  This field can be, in principle, calculated following
the approach recently used in \cite{Carminati,Shchegrov}.
For simplicity,
we shall neglect the thermal emission from the tip itself (which is
assumed to be cold) compared to that of the heated sample. But we do
not need, at this stage, to assume a weak coupling between the tip
and the sample. In particular, in the expressions derived in this
section, the experimental field $\E_{exp}$ is the field emitted by the
sample alone, in the presence of the detecting tip.
Following the approach of \cite{Porto}, based on the reciprocity
theorem of electromagnetism~\cite{reviewSNOM}, an exact relationship
between the signal $\Sig$ and the experimental
field $\E_{exp}$ can be established.
It can be shown that the signal is given by an overlapping integral.

 To proceed, one considers a
fictitious situation in which the sample is removed, and a point
source, represented by a monochromatic current $\jv_{rec}$ oscillating
at frequency $\omega$, is placed at the
position of the detector (see Fig.~\ref{SNOM}(b)). The orientation of 
this
reciprocal source is chosen along the
direction of polarization of the analyzer used in the experimental
situation. The field created around the
tip in this reciprocal situation is denoted by $\E_{rec}$.
Using the reciprocity theorem, the field at the detector
can be written~\cite{Porto}:
\begin{equation}
\E_d(\omega)\cdot\jv_{rec}=\frac{-2i}{\mu_0\om}\int_S
\frac{\partial\E_{rec}(\R,z,\omega)}{\partial z}
\cdot\E_{exp}(\R,z,\omega)d\R \label{reciprocal}
\end{equation}
where the integration is performed in a plane $z=z_0$ between the tip
and the sample and $\R=(x,y)$ are the coordinates along this plane.
Equation (\ref{reciprocal}) connects the field above
the surface $\E_{exp}$ to the field in the detector $\E_d$ along the
direction of the analyzer. Note that the reciprocal field $\E_{rec}$
encodes all the information about the detection system (tip and
collection optics). Reporting the expression of the field at the
detector (\ref{reciprocal}) in (\ref{signal}), one finds the 
expression
for the measured signal:
\begin{equation}
\Sig=\frac{\epsilon_0c}{8\pi^2}\int_S\!\!\int_SH_{ij}(\R,\R',z,\om)
W_{ij}(\R,\R',z,\om)d\R d\R' \ .
\label{signal2}
\end{equation}

Equation (\ref{signal2}) establishes a linear relationship between 
the signal and
the cross-spectral density tensor $W_{ij}$ of the electric field
defined by
\begin{equation}
    \langle E_{exp,i}(\R,z,\omega) 
E_{exp,j}^*(\R^\prime,z,\omega^\prime)
    \rangle = W_{ij}(\R,\R^\prime,z,\omega) \,
    \delta(\omega-\omega^\prime) \ .
    \label{eq:defW}
\end{equation}
The response function $H_{ij}$ only depends on the detection system
(in particular the tip geometry and composition), and is given by
\begin{equation}
H_{ij}(\R,\R',\om)=\frac{\partial\E_{rec,i}(\R,z,\omega)}{\partial z}
\frac{\partial\E_{rec,j}(\R',z,\omega)}{\partial z} \ .
\end{equation}

The cross-spectral density tensor $W_{ij}$ describes the 
electric-field
spatial correlation at a given frequency $\omega$. For the thermal
emission situation considered here, it depends only on the dielectric
constant, on the geometry and on the temperature of the sample.

Equation (\ref{signal2}) is a  general relationship between the
signal and the cross-spectral density tensor. It is non-local and
strongly polarization dependent. This shows
that one do not measure in general a quantity which is proportional to
$W_{kk}(\rv,\rv,\omega)$, and thus to $\rho^E(\rv,\om)$. Nevertheless
Eq.~(\ref{signal2}) suggested
that $\rho^E(\rv,\om)$ can be recovered if the response function $H_{ij}$ is 
localized.
Indeed, in that case the signal is proportional to 
$W_{ij}(\R,\R,z,\omega)$, thus to
$\rho^E(\rv,\om)$. As shown in the next section, a dipole tip (small 
sphere) would exhibit such a response function.

\subsection{Detection of the LDOS by an ideal point-dipole probe}

Let us see what would be measured by an ideal probe consisting of a
single electric dipole described by a polarizability $\alpha(\omega)$.
Note that such a probe was proposed as a model for the uncoated 
dielectric
probe sometimes used in photon scanning tunneling microscopy (PSTM),
and gives good qualitative prediction~\cite{VanLabeke}.
We assume that the thermally emitting medium occupies the
half-space $z<0$, and that the probe is placed at a point $\rv_{t}$.
As in the preceding section, the detector placed in the far field measures
the field intensity at a
given point $\rv_{d}$, through an analyser whose polarization
direction is along the vector $\jv_{rec}$.
In this case, Eq.~(\ref{reciprocal}) simplifies to read
\begin{equation}
    \jv_{rec} \cdot \E_{d} = \alpha(\omega) \,
    \frac{\omega^2}{4\pi c^2} \,
    \frac{\exp(ik|\rv_{d}-\rv_{t}|)}{|\rv_{d}-\rv_{t}|}
    \, \jv_{rec} \cdot \h(\ud) \cdot
    \E_{exp}(\rv_t,\omega)
    \label{eq:dipole}
\end{equation}
where $k=\omega/c$, $\ud=(\rv_{d}-\rv_{t})/|\rv_{d}-\rv_{t}|$ is the
unit vector pointing from the probe towards the detector and
$\h(\ud)=\I-\ud\ud$ is the dyadic
operator which projects a vector on the direction transverse to
$\ud$, $\I$ being the unit dyadic operator. The dyadic $\h(\ud)$
being symmetric, the scalar product in the right-hand side in
Eq.~(\ref{eq:dipole}) can be transformed using the equality
$\jv_{rec} \cdot \h(\ud) \cdot
\E_{exp}(\rv_t,\omega) = \E_{exp}(\rv_t,\omega) \cdot
\h(\ud) \cdot \jv_{rec}$.
Finally, the signal at the detector writes
\begin{equation}
    \langle S \rangle = |\alpha(\omega)|^{2} \frac{\omega^{4}}{4\pi 
c^{4}}
    \, \mathrm{d}\Omega \, \sum_{i,j} A_{i}A_{j}^*
    W_{ij}(\rv_{t},\rv_{t},\omega)
    \label{eq:signal_dip}
\end{equation}
where ${\bf A}=\h(\ud) \cdot \jv_{rec}$
is a vector depending only on the detection conditions
(direction and polarization). Note that if $\jv_{rec}$ is transverse
with respect to the direction $\ud$, which is approximately
the case in many experimental set-ups, then one simply has
${\bf A}=\jv_{rec}$.

Equation (\ref{eq:signal_dip}) shows that with an ideal probe
consisting of a signal dipole (with an isotropic polarizability
$\alpha(\omega)$, one locally measures the cross-spectral density
tensor at the position $\rv_{t}$ of the tip.
Nevertheless, polarization properties of the detection still
exists so that the trace of $W_{ij}$,
and therefore $\rho^E(\rv,\om)$, is not directly measured.
A possibility of measuring the trace would be to measure a signal
$\langle S_{1} \rangle$ in the direction normal to the surface
with an unpolarized detection, and a signal $\langle S_{2} \rangle$
in the direction parallel to the surface, with an analyzer in the
vertical direction. $\langle S_{1} \rangle$ would be a sum of the
two signals obtained with $\jv_{rec}$ along the $x$-direction and
along the $y$-direction. $\langle S_{2} \rangle$ would correspond
to the signal measured with $\jv_{rec}$ along the $z$-direction.
Using Eq.~(\ref{eq:signal_dip}), we see that
the signal $\langle S \rangle = \langle S_{1} \rangle +
\langle S_{2} \rangle$ is proportional to the trace
$W_{kk}(\rv_{t},\rv_{t},\omega)$, and thus to $\rho^E(\rv,\om)$. 
Measuring the thermal spectrum of emission with an apertureless SNOM 
which
probe is dipolar is thus a natural way to achieve the measurement of 
$\rho^E(\rv,\om)$.
Close to the material resonances, i.e in the frequency domain where 
$\rho^E(\rv,\om)$ matches
$\rho(\rv,\om)$, such a near-field thermal emission spectrum gives 
the electromagnetic LDOS.

\subsection{Analogy with scanning (electron) tunneling microscopy}

The result in this section shows that a SNOM measuring the thermally
emitted field with a dipole probe (for example a sphere much smaller
than the existing wavelengths) measures the electromagnetic LDOS
of the sample in the frequency range situated around the resonant 
pulsation. As discussed above, the measured LDOS is that of the
modes which can be excited in the thermal emission process in a cold 
vacuum.
This result was obtained from Eq.~(\ref{reciprocal}) assuming a weak
tip-sample coupling, i.e., the experimental field is assumed to be
the same with or without the tip.

The same result could be obtained starting from the generalized 
Bardeen
formula derived in ref.\cite{Carminati2}. Using this formalism for a 
dipole
probe,
one also ends up with Eq.~(\ref{eq:signal_dip}), which explicitly
shows the linear relationship between the signal and 
$\rho^E(\rv,\om)$.
This derivation is exactly the same as that used in the Tersoff 
and
Haman theory of the STM~\cite{Tersoff}. This theory showed,
in the weak tip-sample coupling limit, that the electron-tunneling
current measured in STM
was proportional to the electronic LDOS of the sample, at the tip
position, and at the Fermi energy. This result, although obtained
under some approximations, was a breakthrough in understanding the
STM signal. In the case of near-field optics, the present discussion,
together with the use of the generalized Bardeen
formula~\cite{Carminati2}, shows that under similar approximations,
a SNOM using an ideal dipole probe and measuring the field thermally
emitted by the sample is the real optical analog to the electron STM.
We believe that this situation provides for SNOM a great potential for
local solid-surface spectroscopy, along the directions opened by STM.

\subsection{Could the LDOS be detected by standard SNOM techniques ?}

Before concluding, we will discuss the ability of standard SNOM 
techniques (by ``standard'' we mean techniques using laser-light 
illumination) to image the electromagnetic LDOS close to a sample. 
Recent experiments~\cite{Chicanne} have shown that an {\it 
illumination-mode} SNOM using metal-coated tips and working in 
transmission produce images which reproduce calculated maps of 
$\rho^E(\rv,\om)$ (which is the adopted definition of the LDOS in 
this experimental work, see also~\cite{Colas}). We shall now show 
that this operating mode bear strong similarites to that 
corresponding to a SNOM working in {\it collection mode}, and 
measuring thermally emitted fields. This will explain why the images 
reproduce  (at least qualitatively) the electric LDOS 
$\rho^E(\rv,\om)$.

Let us first consider a collection-mode technique, in which the 
sample (assumed to be transparent) is illuminated in transmission by 
a monochromatic laser with frequency $\omega$, and the near-field 
light is collected by a local probe. If we assume the illuminating 
light to be spatially incoherent and isotropic in the lower 
half-space (with all incident directions included), then this 
illumination is similar to that produced by thermal fluctuations 
(except that only the modes corresponding to the frequency $\omega$ 
are actually excited). Note that this mode of illumination 
corresponds to that proposed in Ref.~\cite{Carminati3}. This 
similarity, together with the discussion in the precedding paragraph, 
allows to conclude that under these operating conditions, a 
collection-mode SNOM would produce images which closely resemble the 
electric LDOS $\rho^E(\rv,\om)$.

We now turn to the discussion of images produced using an 
illumination-mode SNOM, as that used in Ref.~\cite{Chicanne}. The use 
of the reciprocity theorem allows to derive an equivalence 
between illumination and collection-mode configurations, as shown in 
Ref.~\cite{Mendez}. Starting from the collection-mode instrument 
described above, the reciprocal illumination-mode configuration 
corresponds to a SNOM working in transmission, the light being 
collected by an integrating sphere over all possible transmission 
directions (including those below and above the critical angle). 
Under such conditions, the illumination-mode SNOM produces exactly 
the same image as the collection-mode SNOM using isotropic, spatially 
incoherent and monochromatic illumination. This explains why this 
instrument is able to produce images which closely follow the 
electric LDOS $\rho^E(\rv,\om)$. Finally, note that in 
Ref.~\cite{Chicanne}, the transmitted light is collected above the 
critical angle only, which in principle should be a drawback 
regarding the LDOS imaging. In these experiments, it seems that the 
interpretation of the images as maps of the electric LDOS remains 
nevertheless qualitatively correct, which shows that in this case, 
the main contribution to the LDOS comes from modes with wavevector 
corresponding to propagation directions above the critical angle.

\section{Conclusion}
In this paper, we have introduced a defintion of the 
electromagnetic LDOS $\rho(\rv,\om)$. We have shown that it is fully 
determined by the electric-field Green function, but that in general 
it does not reduce to the trace of its imaginary part 
$\rho^E(\rv,\om)$. We have studied the LDOS variations versus the 
distance to a material surface and have explicitly shown examples in 
which the LDOS deviates from $\rho^E(\rv,\om)$. Nevertheless, we have 
shown that around the material resonances (surface polaritons), the 
near-field LDOS reduces to $\rho^E(\rv,\om)$. Measuring the LDOS with an 
apertureless SNOM using a point-dipole tip should be feasible. The 
principle of the measurement is to record a near-field thermal 
emission spectrum. Under such condition, the instrument behaves as 
the optical analog of the STM, in the weak-coupling regime, which is 
known to measure the electronic LDOS on a metal surface. Finallly, we 
have discussed recent standard SNOM experiments in which the LDOS 
seems to be qualitatively measured. Using general arguments, we have 
discussed  the relevance of such measurements and compared them to 
measurements based on thermal-emission spectroscopy. 

\begin{acknowledgements}
We thank Y. De Wilde, F. Formanek and A.C. Boccara for helpful 
discussions.
\end{acknowledgements}

\appendix

\section{Calculation of the field at the detector for an 
ideal point-dipole probe}

Let $\ev_{rec}(\K)$ and
$\ev_{exp}(\K)$ be the bidimensional Fourier component of 
$\E_{rec}(\rv)$ and $\E_{exp}(\rv)$. In the configuration chosen
in our problem the reciprocal field propagates to the negative $z$ 
whereas the experimental fields propagates to the positive $z$.
Thus
\begin{equation}
   \E_{rec}(\rv)=\int \ev_{rec}(\K) \exp[i(\K.\R-\gamma(\K)z)]d\K
   \label{erecfourier}
    \end{equation}
 \begin{equation}   
       \E_{exp}(\rv)=\int \ev_{exp}(\K) \exp[i(\K.\R+\gamma(\K)z)]d\K
       \label{eexpfourier}
    \end{equation}
where $\gamma(\K)=\sqrt{\om^2/c^2-K^2}$.
Putting Eqs. (\ref{erecfourier}) and (\ref{eexpfourier}) into
Eq.(\ref{reciprocal}) gives
\begin{equation}
\E_d(\omega)\cdot\jv_{rec}=-\frac{8\pi^{2}}{\om\mu_{0}}
\int\gamma(\K)\ev_{rec}(-\K)\cdot\ev_{exp}(\K)d\K
    \end{equation}
$\ev_{exp}(K)$ can be evaluated by calculating the field 
$\E_{exp}(\rv)$.
This last field is the field radiated
by the reciprocal current $\jv_{rec}$ and diffused by the
ideal probe. It can also be seen as the field radiated by the dipole 
induced 
at the position $\rv_t=(\R_t,z_t)$ of the probe. If $\pv$ is the 
dipole induced
at the position of the ideal probe, the reciprocal field at a 
position situated
below the probe writes:
\begin{eqnarray}
\E_{rec}(\rv)=\mu_0\om^2\green(\rv,\rv_t)\cdot\pv \nonumber \\
=\frac{i\mu_0\om^2}{8\pi^2}\int\frac{d^2Ke^{i[\K.(\R-\R_t)+\gamma(z_t-z)]}}{\gamma}
\left[\I-\frac{\kv\kv}{k_0^2}\right]\cdot\pv
\end{eqnarray}
where $k_0^2=\om^2/c^2$.
Comparing this expression and (\ref{erecfourier}),then
\begin{equation}
\ev_{rec}(\K)=\frac{i\mu_0\om^2}{8\pi^2\gamma(\K)}e^{i(-\K.\R_t+\gamma(\K) 
z_t)}\h(\kv^-)\cdot\pv
\end{equation}
where $\kv^-=(\K,-\gamma)$. 
Furthermore, using the fact that $\h(\kv)=\h(-\kv)$ and defining 
$\kv^+=(\K,\gamma)$
\begin{equation}
\ev_{rec}(-\K)=\frac{i\mu_0\om^2}{8\pi^2\gamma}e^{i(\K.\R_t+\gamma 
z_t)}\h(\kv^+)\cdot\pv
\end{equation}
Let us denote $\E(\jv{rec}\rightarrow\rv_t)$ the field
radiated by the reciprocal current in $\rv_t$. The dipole induced 
then writes
$\pv=\alpha(\om)\epsilon_0\E(\jv{rec}\rightarrow\rv_t)$ and
\begin{equation}
\E(\jv{rec}\rightarrow\rv_t)=\frac{i\om\mu_0}{4\pi}\frac{e^{i|\rv_d-\rv_t|}}{|\rv_d-\rv_t|}
\h(\ud)\cdot\jv_{rec}
\end{equation}
Using the fact that for all dyadic $\dyadic$ and for all vector $\av$ 
and $\bv$
\begin{equation}
[\dyadic\cdot\av]\cdot\bv=\av\cdot[\dyadic^T\cdot\bv]
\end{equation}
that $\h$ is a symetric dyadic ($\h=\h^T$),
that $\ev_{exp}(\K)$ is transverse to the direction $\kv^+$ and the
definition of $\E_{exp}(\rv)$.
\begin{equation}
    \jv_{rec} \cdot \E_{d} = \alpha(\omega) \,
    \frac{\omega^2}{4\pi c^2} \,
    \frac{\exp(ik|\rv_{d}-\rv_{t}|)}{|\rv_{d}-\rv_{t}|}
    \, \jv_{rec} \cdot \h(\ud) \cdot
    \E_{exp}(\rv_t,\omega)
\end{equation}

\end{document}